\documentstyle[12pt]{article}

\begin{document}

% The general mathematics macros
% ************************************* MACROS

% bold mu
\def\b #1{\mbox{\footnotesize\boldmath$#1$\normalsize}}

% bold mu barra
\def\bb#1{\mbox{\footnotesize\boldmath$\overline{#1}$\normalsize}}

% ************************************* 0
\title{The Kauffman bracket and the Jones polynomial in quantum gravity}

\author{\small Jorge Griego \\
\small Instituto de F\'{\i}sica, Facultad de Ciencias,\\
\small Trist\'an Narvaja 1674, Montevideo, Uruguay.\\
\small E-mail: griego@fisica.edu.uy}

\date{October 13, 1995}
\maketitle
\abstract{An analysis of the action of the Hamiltonian constraint of
quantum gravity on the Kauffman bracket and Jones knot polynomials is
proposed. It is explicitely shown that the Kauffman bracket is a formal
solution of the Hamiltonian constraint with cosmological constant
($\Lambda$) to third order in $\Lambda$. The calculation is performed in
the extended loop representation of quantum gravity. The analysis makes
use of the analytical expressions of the knot invariants in terms of the
two and three point propagators of the Chern-Simons theory. Some
particularities of the extended loop calculus are considered and the
implications of the results to the case of the conventional loop
representation are discussed.}

\section{Introduction}
In a companion article \cite{Gr} the action of the vacuum Hamiltonian
constraint of quantum gravity on the third coefficient of the Jones knot
polynomial $J_3(\gamma)$ was studied. In that opportunity a formal
(analytical) expression for $H_0\,J_3(\gamma)$ was indirectly obtained
form the fact that the Kauffman bracket is annihilated by $H_{\Lambda}$
to third order in the cosmological constant. The result shows that
$J_3(\gamma)$ can in principle be annihilated by $H_0$ in the loop
representation if the topology of the loop $\gamma$ is appropriately
restricted at the intersecting points.

To complete the analysis developed in \cite{Gr} one has to demonstrate
that the Kauffman bracket is effectually annihilated by the Hamiltonian
constraint with cosmological constant to $O(\Lambda^3)$. The explicit
confirmation of this fact is one of the purposes of this article. We
make use of the extended loop representation of quantum gravity, where
the calculation can be fully developed taking into consideration the
analytical expressions of the knot invariants. In spite that the
regularization and renormalization problems associated
with the formal calculus can be solved using the point splitting method,
we shall limitate the analysis to the formal level for the sake of
simplicity.

Another purpose of this article is to present some features of the
extended loop calculus. There already exists in the literature detailed
references about the extended loop representation and its relationship
with the conventional loop representation of quantum gravity. We do not
give here a self contained introduction on this subject (see
\cite{GaPubook} for a general reference). Nevertheless, the use of
extended loops involves a particular  methodology that is emphasized in
this paper. In particular, it is shown that a systematic of calculation
exists for the constraints in this representation.

The article is organized as follows: in Sect. 2 the relationship between
the Kauffman bracket and the Jones polynomial and their role in quantum
gravity are briefly discussed. Also some mathematical background
necessary for the following sections are introduced. In Set. 3 the
constraints are analyzed from a general point of view in the extended
loop framework. In Sect. 4 we evaluate the Hamiltonian on the third
coefficient of the Jones polynomial. We compute separately the
contributions that appear in the analytical expression of the $J_3$
diffeomorphism invariant. In Sect. 5 the result of $H_0\,J_3$ in terms
of extended loops is analyzed. In Sect. 5.1 we see that the Hamiltonian
does not annihilate the third coefficient of the Jones polynomial for
general extended loops. In Sect. 5.2 a geometrical interpretation in
terms of ordinary loops is developed. The action of the Hamiltonian with
cosmological constant on the Kauffman bracket is considered in Sect. 6.
We confirm that this knot invariant is annihilated by $H_{\Lambda}$ to
$O(\Lambda^3)$. The conclusions are in Sect. 7 and three appendixes with
useful results are added.

\section{Preliminars}
The introduction of the loop representation \cite{RoSm,Ga} has allowed
to make substantial advances in the knowledge of the state of space of
quantum gravity. For one hand, the diffeomorphism invariance of the
theory is automatically implemented by the requirement of the knot
invariance of the loop wavefunctions. By this way, knots and the quantum
states of gravity appears to be related. This relationship is
highlighted by the fact that nonperturbative solutions of the
Wheeler-DeWitt equation can be found in terms of knot invariants
\cite{RoSm,BrGaPu1}. The common feeling about these facts is that knot
theory and general relativity seems to have a profound relationship when
one intends to describe the quantum properties of space-time \cite{Au}.

One of the ways that this relationship is manifested is the following:
for one hand there exists a general link between Chern-Simons and knot
theories \cite{Wi}; on the other the exponential of the Chern-Simons
form constructed with the Ashtekar connections is annihilated by the
constraints of gravity with cosmological constant in the
connection representation \cite{Ko}. As a result, some knot
polynomials could be associated with the solutions of the Hamiltonian
constraint when loops are introduced as the underlying geometrical
structure of quantum gravity. The argumentation is essentially formal
(the loop transform is a formal relationship between the connection and
the loop representations for quantum gravity) but its heuristic use was
proved to be useful in the search of nondegenerate solutions of the
Hamitonian constraint in the conventional loop representation
\cite{BrGaPu2}.

More precisely, the loop transform of the exponential of the
Chern-Simons form is related to the Kauffman bracket knot polynomial
\cite{Wi,Sm,Gumami}. The Kauffman bracket can be expressed as a power
series in the cosmological constant. Each coefficient of the series is a
knot invariant that is, at least formally, promoted as a solution of the
Hamiltonian constraint with cosmological constant in the loop
representation.

There exists a close relationship between the Kauffman bracket and the
Jones polynomials. If $K_{\Lambda}$ and $J_{\Lambda}$ denote these
knot polynomials, one finds that
\begin{equation}
K_{\Lambda}(\gamma)= e^{-\Lambda \varphi_G (\gamma)}\,J_{\Lambda}(\gamma)
\end{equation}
where $\varphi_G (\gamma)$ is the Gauss self-linking number of the loop
$\gamma$. The phase factor $e^{-\Lambda \varphi_G (\gamma)}$ contains
all the framing dependece of the Kauffman bracket invariant and it is by
itself annihilated by the $H_{\Lambda}$ \cite{GaPu,Gr2}. Using this
fact one can write the following expression for the action of the
Hamiltonian on the Kauffman bracket \cite{Gr}

\begin{eqnarray}
&&\hspace{-1cm}
H_{\Lambda} \, K_{\Lambda}(\gamma) = (H_0 + \Lambda \,  det\,q )\,
K_{\Lambda}(\gamma)=\nonumber\\
&&\hspace{-1cm}\sum_{m=2}^{\infty}
\Lambda^m \!\{ H_0\,J_m+\sum_{n=1}^{m-1}
{\textstyle{\frac{(-1)^n}{n!}}}
[H_0 \,(\varphi^{n}_G \,J_{m-n})- n\,det\,q \,(\varphi^n_G \,J_{m-n})]\}
\stackrel{?}{=}0
\label{hkauff}
\end{eqnarray}
where $det\,q$ is the determinant of the three metric and $J_m$ gives
the coefficients of the expansion of the Jones polynomial in terms of
the cosmological constant. The cancellation of this expression is expected
from the formal arguments given before and it has to be checked
explicitely in the loop representation. Notice that for $m=2$ the above
result reduces to (recall that $J_1\equiv 0$)
\begin{equation}
H_0\,J_2 (\gamma)\stackrel{?}{=}0
\end{equation}
By this way the coefficients of the Jones knot polynomial appear to be
related to the solutions of the vacuum Hamiltonian constraint. This fact
and the possibility that a similar result could hold for higher
orders in the cosmological constant was first advanced by Br\"ugmann,
Gambini and Pullin \cite{BrGaPu2}. These authors have confirmed the
annulment of equation (\ref{hkauff}) up to the second order in
$\Lambda$. The next equation reads as follows
\begin{equation}
H_{\Lambda} \, K^{(3)}_{\Lambda} =
\Lambda^3 \{ H_0\,J_3+det\,q \,J_2-H_0 \,(\varphi_G \,J_2)  \}
\stackrel{?}{=}0
\label{hk}
\end{equation}
As it was mentioned, an analysis of this equation was just developed in
\cite{Gr}. In that opportunity a result for $H_0\,J_3(\gamma)$ was
derived from the annulment of (\ref{hk}). Now we are interested to
confirm the fact that $H_{\Lambda} \, K^{(3)}_{\Lambda}(\gamma) =0$.
This means that we have to compute explicitely $H_0\,J_3(\gamma)$. This
calculation confronts with hard computational and regularization
problems in the conventional loop representation. The extended loop
representation offers a new way to solve this problem both at the formal
and the renormalized level.

The relationship between the extended loop and the loop representations
can be formulated in general. The group of loops is a discrete subgroup
of the extended loop group \cite{DiGaGr1}, so any result obtained in the
extended loop framework can be reduced to ordinary loops. In particular
it is demonstrated that the extended version of the constraints of
quantum gravity reduce to the corresponding of the conventional loop
representation once the specialization is performed \cite{DiGaGr2}. We
introduce now some convenient notation related to extended loops. The
elements of the extended loop group are given by infinite strings of
multivector density fields of the form

\begin{equation}
{\bf X} \, =(\, X, \, X^{\, \mu_1}, \, \ldots , \,
X^{\,\mu_1\ldots \mu_n},\, \ldots)
\end{equation}
where $X$ is a real number and a greek index $\mu_i\,:=\,(a_i,\,x_i)$
represents a paired vector index $a_i$ and space point $x_i$. The number
of paired indices defines the {\it rank} of the multivector field. The
constraints have the following expressions in terms of extended loops

\begin{equation}
C_{ax} \psi({\bf R}) = \psi({\cal  F}_{ab}(x) \times
{\bf R}^{(bx)} ) \label{dif}
\end{equation}
\begin{equation}
{\textstyle{1\over2}}H_0 (x) \, \psi({\bf R}) = \psi ({\cal F}_{ab} (x)
\times {\bf R}^{(ax, \,bx)}) \label{ham}
\end{equation}
where $\times$ indicates the extended group product, ${\cal F}_{ab}(x)$
are some elements of the algebra of the group and

\begin{eqnarray}
[{\bf R}^{(bx)}]^{\mu_1\ldots \mu_n} &\equiv&
R^{(bx) \mu_1 \ldots \mu_n} :=
R^{(bx \, \mu_1\ldots \mu_n)_c} \label{runo}\\
\protect[{\bf   R}^{(ax,
\,bx)}]^{\mu_1 \ldots \mu_n} &\equiv& R^{ (ax, \,bx) \mu_1 \ldots \mu_n}
:= \sum_{k=0}^{n} R^{(\, ax \, \mu_1 \ldots \mu_k \, bx \,
\overline{\mu_{k+1} \ldots \mu_n})_c} \label{rdosin}
\end{eqnarray}
We usually call these combinations the one-point-R and the two-point-R
respectively, being an $R$ the following combination of $X$'s

\begin{equation}
R^{\mu_1 \ldots \mu_n}
:= \textstyle{1\over2} [ X^{\mu_1 \ldots \mu_n} + X^{\overline{\mu_1
\ldots \mu_n}}]
\end{equation}
The overline operation is defined as follows

\begin{equation}
X^{\overline{\mu_1 \ldots \mu_n}}:= (-1)^n
X^{\mu_n\ldots \mu_1} \label{overline}
\end{equation}
The subscript $c$ indicates cyclic permutation and in (\ref{rdosin}) the
sequences $\mu_1 \ldots \mu_0$ and $\mu_{n+1} \ldots \mu_n$ for $k=0$
and $k=n$ respectively are assumed to be the null set of indices. Notice
that the diffeomorphism (\ref{dif}) and the Hamiltonian (\ref{ham})
constraints have very similar expressions in the extended loop
representation. The only difference is the object that one
puts into the group product with ${\cal F}_{ab}(x)$. The one- and the
two-point-R have basically different symmetry and regularity properties.
The extended loop wavefunctions are linear in the multivector fields and
they are written in general as follows

\begin{equation}
\psi({\bf   X})   \equiv \psi({\bf   R})=  \sum_{n=0}^{\infty} D_{\mu_1
\ldots \mu_n}   \,    R^{\mu_1 \ldots \mu_n}
\label{wf}
\end{equation}
where the propagators $D_{\mu_1 \ldots \mu_n}$ satisfy a set of
identities of the Mandelstam type \cite{DiGaGr2} and a generalized
convention of sum are assumed for the repeated paired (greek) indices.
Usually we refer the $_{\mu_i}$ indices of the propagators as {\it
covariant} whereas the $^{\mu_i}$ indices of the multivector fields will
be called {\it contravariant}. This terminology is suggested from the
behavoir of these objects under general coordinate transformations
\cite{GaLe,DiGaGr1}. Using (\ref{wf}) we get from (\ref{ham}) the
following general expression for the action of the Hamiltonian on the
extended loop wavefunctions

\begin{equation}
\textstyle{1\over2}H_0 (x)  \psi({\bf R}) =
\sum_{n=0}^{\infty} D_{\mu_1 \ldots \mu_n}[
{\cal F}_{ab}^{\mu_1} (x)\,   R^{(ax,\,bx)\mu_2 \ldots \mu_n}+
{\cal F}_{ab}^{\mu_1 \mu_2} (x)\,   R^{(ax,\,bx)\mu_3 \ldots \mu_n}]
\label{ham2}
\end{equation}
where we have explicitely used the fact that ${\cal F}_{ab}(x)$ has only
two nonvanishing components ${\cal F}_{ab}^{\mu_1} (x)$ and ${\cal
F}_{ab}^{\mu_1 \mu_2} (x)$, given by

\begin{eqnarray}
&& {\cal F}_{ab}{^{\mu_1}}(x)   =
\delta_{a\,b}^{a_1   d} \; \partial_d
\, \delta(x_1 - x) \label{f1}\\
&&  {\cal  F}_{ab}{^{\mu_1 \mu_2}}   (x)     =
\delta_{a\,b}^{a_1 a_2} \; \delta(x_1 - x) \, \delta(x_2 - x)
\label{f2}
\end{eqnarray}
The propagators are cyclic: $D_{\mu_1 \ldots \mu_n}\equiv D_{(\mu_1
\ldots \mu_n)_c}$.  This means that the indices of the propagator that
are contracted with the ${\cal F}$'s really lie in any position of the
$D$'s. It is a remarkable fact that for all the (known) wavefunctions of
quantum gravity, the propagators $D_{\mu_1 \ldots \mu_n}$ are expressed
completely in terms of  the two and three point propagators of the
Chern-Simons theory $g_{\mu_1 \mu_2}$ and $h_{\mu_1 \mu_2 \mu_3}$. We
write the Chern-Simons propagators in the following way

\begin{eqnarray}
g_{\mu_1 \mu_2}   &:=& \epsilon_{a_1 a_2 \,k}\; \phi^{\,kx_1}_{\,x_2}
\label{metrica}\\
h_{\mu_1 \mu_2 \mu_3} &:=& \epsilon^{\alpha_1 \alpha_2 \alpha_3} \,
g_{\mu_1 \alpha_1} \, g_{\mu_2 \alpha_2}\, g_{\mu_3 \alpha_3}
\label{h}
\end{eqnarray}
with
\begin{equation}
\phi^{\,kx_1}_{\,x_2}:= - \frac{1}{4\pi}
\frac{(x_1 -x_2)^k}{\mid x_1 -x_2\mid^3}= -\frac{\partial^{\,k}}
{\nabla^{2}} \, \delta(x_1 -x_2)
\label{phi}
\end{equation}
and
\begin{equation}
\epsilon^{\alpha_1 \alpha_2 \alpha_3}:= \epsilon^{c_1 c_2 c_3}
\delta(z_1 -z_2)\,\delta(z_1 -z_3)
\label{epsilon}
\end{equation}
In the above equations we have used a combined notation for the indices
$\mu_i=(a_i,\,x_i)$ and $\alpha_k=(c_k,\,z_k)$. Also notice that in
(\ref{ham2}) the greek indices of ${\cal F}_{ab}^{\mu_1} (x)$ and ${\cal
F}_{ab}^{\mu_1 \mu_2} (x)$ are contracted with $g$'s and $h$'s (the
building blocks of the $D$'s) in different arrangement of the covariant
indices. In the next section we develop the result of these contractions
for a generic case.

\section{The ${\cal F}_{ab}(x)$'s on the Chern-Simons \protect\\
propagators}
The components of ${\cal F}_{ab}(x)$ are distributional objects of rank
one and two and their action on the propagators $g_{\cdot \cdot}$ and
$h_{\cdot \cdot \cdot}$ can be worked explicitely at the formal level.
According to (\ref{f2}) the component of rank two includes only
(discrete and continuous) delta functions, so its action on the
propagators is straightforward

\begin{eqnarray}
&&{\cal F}_{ab}{^{\mu_1 \mu_2}}(x) \, g_{ \mu_1 \mu_3 }
g_{\mu_2 \mu_4}  =   g_{\mu_3  [ ax } \, g_{\, bx] \, \mu_4 }
\label{F2g}\\
&&{\cal F}_{ab}{^{\mu_1 \mu_2}}(x) \, h_{ \mu_1 \mu_2 \mu_3}
 =   2\, h_{ax \,  bx \, \mu_3} \label{F2h}
\end{eqnarray}
Notice that in the case that the two contravariant indices of ${\cal
F}_{ab}{^{\mu_1 \mu_2}}(x)$ are joined to the same $g$, the
(divergent) contribution vanishes by symmetry. In order to take into
account the derivative that appears in (\ref{f1}) it is convenient to
write ${\cal F}_{ab}^{\mu_1} (x)$ in terms of the inverse of the two
point propagator, given by
\begin{equation}
g^{\mu_1 \mu_2} = \epsilon^{a_1 a_2 \,k}\;\partial_k \,\delta (x_1 - x_2)
\label{invmetrica}
\end{equation}
{}From (\ref{f1}) we have
\begin{equation}
{\cal F}_{ab}{^{\mu_1}}(x) =- \epsilon_{abc}\epsilon^{c\,a_1 d}\;
\partial_d\, \delta(x - x_1)=- \epsilon_{abc}\,g^{cx\,\mu_1}
\end{equation}
The action of ${\cal F}_{ab}{^{\mu_1}}(x)$ over the two point propagator
is simply given by

\begin{equation}
{\cal F}_{ab}{^{\mu_1}}(x) \, g_{\mu_1 \mu_2} =
- \epsilon_{abc}\,g^{cx\,\mu_1}\, g_{\mu_1 \mu_2} =
- \epsilon_{abc} \delta{^{\phantom{T} cx}_{\mbox{\scriptsize{T}}\phantom{ax}
\mu_2}}
\label{f1g}
\end{equation}
where $\delta{^{\phantom{T} cx}_{\mbox{\scriptsize{T}}\phantom{ax}
\mu_2}}$ is the projector on the space of transverse (divergence free)
multivector density fields \cite{DiGaGr1}, given by
\begin{equation}
\delta{^{\phantom{T} cx}_{\mbox{\scriptsize{T}}\phantom{ax}
\mu_2}}= \delta^{c}_{a_2}\delta(x-x_2)-\phi^{\,cx}_{\,x_2,\,a_2}
\label{deltat}
\end{equation}
Notice now in which way ${\cal F}_{ab}{^{\mu_1}}(x)$ operates on a
generic term of a wavefunction when it is contracted with a two point
propagator:

\begin{eqnarray}
\hspace{-1cm}
{\cal F}_{ab}{^{\mu_1}}(x) \, g_{\mu_1 \alpha_k}D_{\ldots \ldots}
R^{\ldots\, \alpha_k \ldots}\!\! &=&\!\!- \epsilon_{abc}
\delta{^{\phantom{T} cx}_{\mbox{\scriptsize{T}}\phantom{ax}
\alpha_k}}\,D_{\ldots\ldots}R^{\ldots\, \alpha_k \ldots}
\nonumber\\
&=&\!\!- \epsilon_{abc}D_{\ldots \ldots}
[R^{\ldots \,cx\, \ldots}
-\phi^{\,cx}_{\,z_k,\,c_k}\,R^{\ldots\, \alpha_k \ldots}]
\nonumber\\
&=&\!\!- \epsilon_{abc}D_{\ldots \ldots}
[R^{\ldots \,cx\, \ldots}+\phi^{\,cx}_{\,z_k}\,
\partial_{\alpha_k}R^{\ldots\, \alpha_k \ldots}]
\label{prueba}
\end{eqnarray}
where only the indices of interest were written and in the last step the
second term was integrated by parts with respect to the $z_k$ variable
($\partial_{\alpha_k}\equiv\partial_{c_k z_k}\equiv\partial/\partial
z_k^{c_k}$). The integration by parts induced by the longitudinal
projector $\phi^{\,cx}_{\,z_k,\,c_k}$ produces the divergence of the
multivectors with respect to the $\alpha_k$-entry. The divergence taking
with respect to any index of an extended loop of rank $n$ generates a
contribution of rank $n-1$ in the following way

\begin{equation}
\partial_{\alpha_k}R^{\ldots\, \alpha_k \ldots}=
[ \delta(z_k-z_{k-1}) - \delta(z_k-z_{k+1}) ]
R^{\cdot \cdot \alpha_{k-1}\alpha_{k+1} \cdot \cdot}
\label{dc}
\end{equation}
where $\alpha_{k-1}$ and $\alpha_{k+1}$ are the neighbors of $\alpha_k$
in the sequence of indices of $R$ (for $k=1$ and $k=n$ we have
$z_0\,=\,z_{n+1}\,\equiv o$, with $o$ a reference spatial point). The above
property is the ``differential constraint" of the elements of the
extended loop group \cite{DiGaGr1}. Applying (\ref{dc}) to
(\ref{prueba}) we get

\begin{equation}
{\cal F}_{ab}{^{\mu_1}}(x)g_{\mu_1 \alpha_k}D_{\ldots \ldots}
R^{\ldots\, \alpha_k \ldots}\!=\!
- \epsilon_{abc}D_{\ldots \ldots}[R^{\ldots \,cx\, \ldots}
+(\phi^{\,cx}_{\,z_{k-1}}\!-\phi^{\,cx}_{\,z_{k+1}}\!)
R^{\ldots\, \alpha_{k-1}\alpha_{k+1} \ldots}]
\label{prueba2}
\end{equation}
This equation exhibits the effect of ${\cal F}_{ab}{^{\mu_1}}(x)$ on the
wavefunction when the $\mu_1$ index lies on a two point propagator. We
see that for any term of rank  $n$ of the expansion (\ref{wf}), a term
of rank $n-1$ is always generated by means of the differential
constraint (\ref{dc}). These type of contributions are characterized by
the appearance of a difference of $\phi$ functions. Also notice that in
the case that ${\cal F}_{ab}{^{\mu_1}}(x)$ acts on a three point
propagator $h$ instead of a two point $g$, the only difference with the
previous result is that the $\alpha_k$ index is now linked with others
$g$'s throught $\epsilon^{\alpha_k \cdot \cdot}$ according to (\ref{h}).
In effect, we have in this case

\begin{equation}
{\cal F}_{ab}{^{\mu_1}}(x) \, h_{\mu_1 \pi_i \pi_j} =-\epsilon_{abc}
\delta{^{\phantom{T} cx}_{\mbox{\scriptsize{T}}\phantom{ax}
\alpha_k}}\epsilon^{\alpha_k \alpha_l \alpha_m}\,
g_{\pi_i \alpha_l}\, g_{\pi_j \alpha_m}
\label{f1h2}
\end{equation}
In the appendix I it is shown that
\begin{equation}
\delta{^{\phantom{T} cx}_{\mbox{\scriptsize{T}}\phantom{ax}
\alpha_k}}\epsilon^{\alpha_k \alpha_l \alpha_m}=
\epsilon^{cx\, \alpha_l \alpha_m}+(\phi^{\,cx}_{\,z_l}
-\phi^{\,cx}_{\,z_m})g^{\alpha_l \alpha_m}
\end{equation}
Using this fact equation (\ref{f1h2}) can be put in the form

\begin{equation}
{\cal F}_{ab}{^{\mu_1}}(x) \, h_{\mu_1 \pi_i \pi_j} =
-g_{\pi_i  [\,ax} g_{\, bx] \, \pi_j} +
\epsilon_{abc}\,\phi^{\,cx}_{\,z}
\delta{^{\phantom{T} dz}_{\mbox{\scriptsize{T}}\phantom{dz}
[\,\pi_i}} \,g_{\pi_j]\,dz}
\label{f1h3}
\end{equation}
where in the last term an integration in $z$ is assumed. When the
$\pi\equiv(e,y)$ indices are contracted with contravariant indices of
multivector fields, the differential constraint (\ref{dc}) will be
induced by the longitudinal projectors $\phi^{\,dz}_{\,y_i,\,e_i}$ and
$\phi^{\,dz}_{\,y_j,\,e_j}$. The following result is obtained for this
case

\begin{eqnarray}
&&\hspace{-1cm}{\cal F}_{ab}{^{\mu_1}}(x) \, h_{\mu_1 \pi_i
\pi_j}D_{\ldots \ldots \ldots}R^{\ldots\, \pi_i \ldots \pi_j
\ldots}= \nonumber\\
&&\{-g_{\pi_i  [\,ax} g_{\, bx] \, \pi_j}
+\epsilon_{abc}\,(\phi^{\,cx}_{\,y_i}\,
-\phi^{\,cx}_{\,y_j})g_{\pi_i \pi_j}\} D_{\ldots \ldots
\ldots}R^{\ldots\, \pi_i \ldots \pi_j \ldots}\nonumber\\
&&+\epsilon_{abc}\,\phi^{\,cx}_{\,z}
(\phi^{\,dz}_{\,y_{j-1}}-\phi^{\,dz}_{\,y_{j+1}})
g_{\pi_i \,dz} D_{\ldots \ldots\ldots}
R^{\ldots\, \pi_i \ldots \pi_{j-1}\pi_{j+1} \ldots}\nonumber\\
&&+\epsilon_{abc}\,\phi^{\,cx}_{\,z}
(\phi^{\,dz}_{\,y_{i+1}}-\phi^{\,dz}_{\,y_{i-1}})
g_{\pi_j \,dz} D_{\ldots\ldots\ldots}
R^{\ldots\, \pi_{i-1}\pi_{i+1} \ldots \pi_j \ldots}
\label{prueba3}
\end{eqnarray}
We see that a larger number of $\phi$'s appear when the
$\mu_1$ index of the propagator belongs to an $h$.
This is a general result: suppose that the index $\pi_j$ of the $h$ is
connected with others $g$'s throught an $\epsilon^{\pi_j \cdot
\cdot}$ (instead to be contracted with a contravariant multivector index
as in (\ref{prueba3})); then a greater number of integrations by parts
involving longer chains of $\phi$'s would appear in the final result.
Notice that this is the case for the term of lower rank of $J_3({\bf
R})$ (see equation (\ref{j3}) below). In the appendix II we shall
prove that

\begin{eqnarray}
\lefteqn{ {\cal F}_{ab}^{\mu_1} (x)h_{\mu_1 \mu_2 \alpha}
g^{\alpha\beta} h_{\mu_3 \mu_4 \beta} = }
\nonumber\\
&&- g_{\mu_2  [\,ax} h_{\,bx] \,\mu_3 \mu_4 }
+\epsilon_{abc}\,\phi^{\,cx}_{\,x_{2}}\,h_{\mu_2 \mu_3\mu_4}
-\epsilon_{abc}\,\phi^{\,cx}_{\,z}\,\epsilon^{dz\,\pi_1\pi_2}
g_{\mu_2  \,dz}g_{\mu_3\pi_1} g_{\mu_4  \pi_2}
\nonumber\\
&&-\epsilon_{abc}\,\phi^{\,cx}_{\,z}\,
\phi^{\,dz}_{\,x_2,\,a_2}\,h_{dz\,\mu_3 \mu_4}
+\epsilon_{abc}\,\phi^{\,cx}_{\,z}\,
(\phi^{\,dz}_{\,x_3}\,-\phi^{\,dz}_{\,x_4})
g_{\mu_2  \,dz}g_{\mu_3 \mu_4}
\nonumber\\
&&+\epsilon_{abc}\,\phi^{\,cx}_{\,z}\,
\phi^{\,dz}_{\,y}\,
(\phi^{\,ey}_{\,x_4,\,a_4}g_{\mu_3\,ey}-
\phi^{\,ey}_{\,x_3,\,a_3}g_{\mu_4\,ey})g_{\mu_2  \,dz}
\label{F1hh}
\end{eqnarray}
In this case we will have three integrations by parts once the
free covariant indices of the above expression are contracted with
contravariant greek indices of multivector fields. The above discussion
puts in evidence that a systematization exists for the calculation in
the extended loop representation of quantum gravity. It is a remarkable
fact that the systematic (that is based essentially on the properties of
${\cal F}_{ab}(x)$ and the two and three point propagators of the
Chern-Simons theory) is the {\it same} for {\it both} the Hamiltonian
and diffeomorphism constraints. This last property is a consequence of
the fact that the combinations $R^{(bx) \mu_1 \ldots \mu_n}$ and $R^{
(ax, \,bx) \mu_1 \ldots \mu_n}$ satisfies the differential constraint
with respect to the $\mu$ indices in the two cases. Another aspects of
the methodology of the extended loop calculus will be pointed out in the
following sections.

\section{The Hamiltonian on $\bf{J_3}$}

The analytical expression of the $J_3$ diffeomorphism invariant is
\cite{DiGr}
\begin{eqnarray}
J_3 ({\bf R}) &=&
-6\{(2g_{\mu_1\mu_4}g_{\mu_2\mu_5}g_{\mu_3\mu_6} +
{\textstyle{1 \over 2}}
g_{(\mu_1\mu_3}g_{\mu_2\mu_5}g_{\mu_4\mu_6)_c}) \,
R^{\mu_1\mu_2\mu_3\mu_4\mu_5\mu_6}
\nonumber \\
&& \hspace{1.7cm}+ g_{(\mu_1\mu_3}h_{\mu_2\mu_4\mu_5)_c} \,
R^{\mu_1\mu_2\mu_3\mu_4\mu_5} \nonumber \\
&&\hspace{0.9cm}+(h_{\mu_1\mu_2\alpha} g^{\alpha\beta} h_{\mu_3\mu_4\beta} -
h_{\mu_1\mu_4\alpha} g^{\alpha\beta} h_{\mu_2\mu_3\beta}) \,
R^{\mu_1\mu_2\mu_3\mu_4}\}
\label{j3}
\end{eqnarray}
In order to evaluate the Hamiltonian on $J_3$ it is
convenient to consider the contributions of the different ranks for
separate. In the next subsections we shall develop the partial results
for the different ranks.

\subsection{$H_0\,(g_{\cdot \cdot}g_{\cdot \cdot}g_{\cdot \cdot}
R^{\cdot \cdot \cdot \cdot \cdot \cdot })$}

Let us start by considering the action of the rank one component of ${\cal
F}_{ab}(x)$ on the terms with three two point propagators. We have
\begin{eqnarray}
&&{\cal F}_{ab}{^{\cdot}}\,(x)\,g_{\cdot \cdot}g_{\cdot \cdot}
g_{\cdot \cdot}\,R^{(ax,\,bx)\cdot\cdot\cdot\cdot\cdot}=
{\cal F}_{ab}{^{\mu_1}}(x)(2g_{\mu_1\mu_4}g_{\mu_2\mu_5}g_{\mu_3\mu_6}
+ g_{\mu_1\mu_3}g_{\mu_2\mu_5}g_{\mu_4\mu_6}
\nonumber\\
&&+g_{\mu_1\mu_5}g_{\mu_2\mu_4}g_{\mu_3\mu_6}
+g_{\mu_1\mu_4}g_{\mu_2\mu_6}g_{\mu_3\mu_5})
R^{(ax,\,bx)\mu_2 \mu_3 \mu_4 \mu_5\mu_6}
\label{f1ggg}
\end{eqnarray}
This expression can be simplified using the following property of the
two-point-R:
\begin{equation}
R^{(ax,\,bx)\overline{\mu_1 \ldots \mu_n}}=
R^{(bx,\,ax)\mu_1 \ldots \mu_n}
\label{Rsim}
\end{equation}
This property follows directly form the fact that the $R$'s are
{\it even} with respect to the overline operation. Notice now that
\begin{eqnarray}
\lefteqn{ {\cal F}_{ab}{^{\mu_1}}(x)
g_{\mu_1\mu_5}g_{\mu_2\mu_4}g_{\mu_3\mu_6}
R^{(ax,\,bx)\mu_2 \mu_3 \mu_4 \mu_5\mu_6} }
\nonumber\\
&&=-{\cal F}_{ab}{^{\mu_1}}(x)
g_{\mu_1\mu_5}g_{\mu_2\mu_4}g_{\mu_3\mu_6}
R^{(bx,\,ax)\mu_6 \mu_5 \mu_4 \mu_3\mu_2}
\nonumber\\
&&\equiv{\cal F}_{ab}{^{\mu_1}}(x)
g_{\mu_1\mu_3}g_{\mu_2\mu_5}g_{\mu_4\mu_6}
R^{(ax,\,bx)\mu_2 \mu_3 \mu_4 \mu_5\mu_6}
\end{eqnarray}
This kind of symmetry operation will be used repeatedly along the calculation.
{}From (\ref{f1g}) and (\ref{deltat}) we get
\begin{eqnarray}
\hspace{-0.33cm}
\lefteqn{ {\cal F}_{ab}{^{\cdot}}\,(x)\,g_{\cdot \cdot}g_{\cdot \cdot}
g_{\cdot \cdot}\,R^{(ax,\,bx)\cdot\cdot\cdot\cdot\cdot} }
\nonumber\\
&&\hspace{-0.33cm}=-\epsilon_{abc}\{
\delta{^{\phantom{T} cx}_{\mbox{\scriptsize{T}}\phantom{ax}\mu_4}}
(2g_{\mu_2\mu_5}g_{\mu_3\mu_6}+g_{\mu_2\mu_6}g_{\mu_3\mu_5})
\nonumber\\
&&\hspace{-0.33cm}
+2\delta{^{\phantom{T} cx}_{\mbox{\scriptsize{T}}\phantom{ax}\mu_3}}
g_{\mu_2\mu_5}g_{\mu_4\mu_6}\}
R^{(ax,\,bx)\mu_2 \mu_3 \mu_4 \mu_5\mu_6}
\nonumber\\
&&\hspace{-0.33cm}=-2\epsilon_{abc}g_{\mu_1\mu_3}g_{\mu_2\mu_4}
(R^{(ax,\,bx)\mu_1\,cx\,\mu_2\mu_3\mu_4}
+R^{(ax,\,bx)\mu_1\mu_2\,cx\,\mu_3\mu_4})
\nonumber\\
&&\hspace{-0.33cm}-\epsilon_{abc}g_{\mu_1\mu_4}g_{\mu_2\mu_3}
R^{(ax,\,bx)\mu_1\mu_2\,cx\,\mu_3\mu_4}
\nonumber\\
&&\hspace{-0.33cm}
-\epsilon_{abc}\{2(\phi^{\,cx}_{\,x_1}\!-\!\phi^{\,cx}_{\,x_3}\!)
g_{\mu_1\mu_3}g_{\mu_2\mu_4}\!+\!
(\phi^{\,cx}_{\,x_2}\!-\!\phi^{\,cx}_{\,x_3}\!)g_{\mu_1\mu_4}g_{\mu_2\mu_3}
\}R^{(ax,\,bx) \mu_1 \mu_2 \mu_3 \mu_4}
\label{r7}
\end{eqnarray}
For the component of rank two one obtains
\begin{eqnarray}
\lefteqn{ {\cal F}_{ab}{^{\cdot\cdot}}(x)g_{\cdot\cdot}g_{\cdot\cdot}
g_{\cdot \cdot}R^{(ax,\,bx)\cdot\cdot\cdot\cdot}= }
\nonumber\\
&&\{2g_{\mu_1\mu_3}g_{\mu_2  [ ax } \, g_{\, bx] \, \mu_4 }
+g_{\mu_1\mu_4}g_{\mu_2  [ ax } \, g_{\, bx] \, \mu_3 }\}
R^{(ax,\,bx) \mu_1 \mu_2 \mu_3 \mu_4}
\end{eqnarray}
The partial result for the rank six of $J_3$ is
\begin{eqnarray}
&&\lefteqn{
\textstyle{1\over12}H_0\,(g_{\cdot \cdot}g_{\cdot \cdot}g_{\cdot \cdot}
R^{\cdot \cdot \cdot \cdot \cdot \cdot })=
\epsilon_{abc}g_{\mu_1\mu_4}g_{\mu_2\mu_3}
R^{(ax,\,bx)\mu_1\mu_2\,cx\,\mu_3\mu_4} }
\nonumber\\
&&+2\epsilon_{abc}g_{\mu_1\mu_3}g_{\mu_2\mu_4}
[R^{(ax,\,bx)\mu_1\,cx\,\mu_2\mu_3\mu_4}
+R^{(ax,\,bx)\mu_1\mu_2\,cx\,\mu_3\mu_4}]
\nonumber\\
&&+\{-2g_{\mu_1\mu_3}g_{\mu_2  [ ax } \, g_{\, bx] \, \mu_4 }
-g_{\mu_1\mu_4}g_{\mu_2  [ ax } \, g_{\, bx] \, \mu_3 }
+\epsilon_{abc}\{2(\phi^{\,cx}_{\,x_1}\!-\!\phi^{\,cx}_{\,x_3}\!)
g_{\mu_1\mu_3}g_{\mu_2\mu_4}
\nonumber\\
&&+(\phi^{\,cx}_{\,x_2}\!-\!\phi^{\,cx}_{\,x_3}\!)g_{\mu_1\mu_4}g_{\mu_2\mu_3}
\}R^{(ax,\,bx) \mu_1 \mu_2 \mu_3 \mu_4}
\label{Hr6}
\end{eqnarray}
We get three types of terms: one of rank seven and two of rank six. The
terms of rank seven have the multivector fields with three spatial
indices evaluated at the point $x$. The contribution of rank six has two
different sources: for one hand the terms generated by the application
of the differential constraint (\ref{dc}) on the multivector fields of
rank seven and  on the other the terms generated by the action of ${\cal
F}_{ab}^{\mu_1\mu_2}(x)$ on the propagators.

\subsection{$H_0\,(g_{\cdot \cdot}h_{\cdot \cdot \cdot}
R^{\cdot \cdot \cdot \cdot \cdot })$}

We have in this case
\begin{eqnarray}
\lefteqn{ {\cal F}_{ab}{^{\cdot}}\,(x)\,
g_{\cdot \cdot}h_{\cdot \cdot \cdot}\,R^{(ax,\,bx)\cdot\cdot\cdot\cdot} }
\nonumber\\
&&={\cal F}_{ab}{^{\mu_1}}(x)(2g_{\mu_1\mu_3}h_{\mu_2\mu_4\mu_5} +
2g_{\mu_2\mu_4}h_{\mu_1\mu_3\mu_5}+g_{\mu_2\mu_5}h_{\mu_1\mu_3\mu_4})
R^{(ax,\,bx)\mu_2\mu_3\mu_4\mu_5}
\nonumber\\
&&=\{-2\epsilon_{abc}
\delta{^{\phantom{T} cx}_{\mbox{\scriptsize{T}}\phantom{ax}\mu_3}}
h_{\mu_2\mu_4\mu_5}
-2g_{\mu_2\mu_4}g_{\mu_3[ax}\,g_{\,bx]\,\mu_5}
+2\epsilon_{abc}g_{\mu_2\mu_4}\phi^{\,cx}_{\,z}
\delta{^{\phantom{T} dz}_{\mbox{\scriptsize{T}}\phantom{dz}
[\,\mu_3}} \,g_{\mu_5]\,dz}
\nonumber\\
&&-g_{\mu_2\mu_5}g_{\mu_3[ax}\,g_{\,bx]\,\mu_4}
+\epsilon_{abc}g_{\mu_2\mu_5}\phi^{\,cx}_{\,z}
\delta{^{\phantom{T} dz}_{\mbox{\scriptsize{T}}\phantom{dz}
[\,\mu_3}} \,g_{\mu_4]\,dz}
\}R^{(ax,\,bx)\mu_2\mu_3\mu_4\mu_5}
\label{f1gh}
\end{eqnarray}
where in the last step we have used (\ref{f1g}) and (\ref{f1h3}).
Introducing now (\ref{deltat}) and performing the integration by parts
indicated by the longitudinal projectors one obtains after a few direct
manipulations
\begin{eqnarray}
\lefteqn{ {\cal F}_{ab}{^{\cdot}}\,(x)\,
g_{\cdot \cdot}h_{\cdot \cdot \cdot}\,R^{(ax,\,bx)\cdot\cdot\cdot\cdot}=
-2\epsilon_{abc}h_{\mu_1\mu_2\mu_3}
R^{(ax,\,bx)\mu_1 \,cx\, \mu_2 \mu_3} }
\nonumber\\
&&+\{-2g_{\mu_1\mu_3}g_{\mu_2[ax}\,g_{\,bx]\,\mu_4}
-g_{\mu_1\mu_4}g_{\mu_2[ax}\,g_{\,bx]\,\mu_3}
+2\epsilon_{abc}(\phi^{\,cx}_{\,x_1}\!-\!\phi^{\,cx}_{\,x_3}\!)
g_{\mu_1\mu_3}g_{\mu_2\mu_4}
\nonumber\\
&&+\epsilon_{abc}(\phi^{\,cx}_{\,x_2}\!-\!\phi^{\,cx}_{\,x_3}\!)
g_{\mu_1\mu_4}g_{\mu_2\mu_3}\}R^{(ax,\,bx)\mu_1\mu_2\mu_3\mu_4}
\nonumber\\
&&+2\{-\epsilon_{abc}\phi^{\,cx}_{\,x_1}h_{\mu_1\mu_2\mu_3}
+\epsilon_{abc}\phi^{\,cx}_{\,z}
(\phi^{\,dz}_{\,x_3}-\phi^{\,dz}_{\,x_2})g_{\mu_1 \,dz}g_{\mu_2\mu_3}
\nonumber\\
&&+\epsilon_{abc}\phi^{\,cx}_{\,z}
\phi^{\,dz}_{\,x}g_{\mu_2 \,dz}g_{\mu_1\mu_3}\}
R^{(ax,\,bx)\mu_1 \mu_2 \mu_3}
\end{eqnarray}
The component of rank two of ${\cal F}_{ab}(x)$ gives the
following contribution
\begin{equation}
{\cal F}_{ab}{^{\cdot\cdot}}(x)g_{\cdot\cdot}h_{\cdot\cdot\cdot}
R^{(ax,\,bx)\cdot\cdot\cdot}=
2\{g_{\mu_1[\,ax}h_{bx \,]\,\mu_2\mu_3}+g_{\mu_1\mu_3}h_{ax\,bx\,\mu_2}\}
R^{(ax,\,bx) \mu_1 \mu_2 \mu_3}
\end{equation}
The partial result for the rank five of $J_3$ is
\begin{eqnarray}
&&\textstyle{1\over12}H_0\,(g_{\cdot \cdot}h_{\cdot \cdot \cdot}
R^{\cdot \cdot \cdot \cdot \cdot })=2\epsilon_{abc}h_{\mu_1\mu_2\mu_3}
R^{(ax,\,bx)\mu_1 \,cx\, \mu_2 \mu_3}
\nonumber\\
&&-\{-2g_{\mu_1\mu_3}g_{\mu_2[ax}\,g_{\,bx]\,\mu_4}
-g_{\mu_1\mu_4}g_{\mu_2[ax}\,g_{\,bx]\,\mu_3}
+2\epsilon_{abc}(\phi^{\,cx}_{\,x_1}\!-\!\phi^{\,cx}_{\,x_3}\!)
g_{\mu_1\mu_3}g_{\mu_2\mu_4}
\nonumber\\
&&+\epsilon_{abc}(\phi^{\,cx}_{\,x_2}\!-\!\phi^{\,cx}_{\,x_3}\!)
g_{\mu_1\mu_4}g_{\mu_2\mu_3}\}R^{(ax,\,bx)\mu_1\mu_2\mu_3\mu_4}
\nonumber\\
&&-2\{g_{\mu_1[\,ax}h_{bx \,]\,\mu_2\mu_3}
-\epsilon_{abc}\phi^{\,cx}_{\,x_1}h_{\mu_1\mu_2\mu_3}
\nonumber\\
&&+\epsilon_{abc}\phi^{\,cx}_{\,z}
(\phi^{\,dz}_{\,x_3}-\phi^{\,dz}_{\,x_2})g_{\mu_1 \,dz}g_{\mu_2\mu_3}
\}R^{(ax,\,bx)\mu_1 \mu_2 \mu_3}
\label{Hr5}
\end{eqnarray}
where we have used (\ref{enap3}). Three different sort of terms appear
in this case: two of rank six and one of rank five. One of the
contributions of rank six have the multivector fields with three spatial
indices evaluated at the point $x$ (like the rank seven of (\ref{r7})).
The other {\it cancels} exactly the contribution of rank six that comes
from $\textstyle{1\over12}H_0\,(g_{\cdot \cdot}g_{\cdot \cdot}g_{\cdot
\cdot} R^{\cdot \cdot \cdot \cdot \cdot \cdot })$. Moreover, we will see
in the next subsection that the rank five of (\ref{Hr5}) is canceled by
terms that appears when the Hamiltonian acts on the term of rank four of
$J_3$.

The above discussion exhibits another property of the constraints in the
extended loop representation. A chain mechanism of cancellations appears
when the constraints act on a wavefunction. This mechanism links
intimately the successive ranks of the wavefunction. One can summarize
this mechanism in the following way for the case of the Hamiltonian: the
Hamiltonian acting on the rank $n$ of the wavefunction generates a
contribution of rank $n+1$ and other of rank $n$. In the (remnant)
contribution of rank $n+1$ the multivector fields have always three
spatial points evaluated at $x$ (they are of the general form
$R^{(ax,\,bx)\ldots\,cx\,\ldots}$). The contribution of rank $n$ is
canceled by terms that are generated when the operator acts on the rank
$n-1$ of the wavefunction. We will see in the next subsection that the
last term of $J_3$ closes the chain in a consistent way. It is important
to remark that exactly the same cancellations take place in the case of
the diffeomorphism operator.

\subsection{$H_0\,(h_{\cdot \cdot \star}g^{\star \star}h_{\star \cdot
\cdot} R^{\cdot \cdot \cdot \cdot })$}

The action of the Hamiltonian on the term of rank four of $J_3$ is given
by the following expression
\begin{eqnarray}
\lefteqn{-\textstyle{1\over12}
H_0\,(h_{\cdot \cdot \star}g^{\star \star}h_{\star \cdot
\cdot}R^{\cdot \cdot \cdot \cdot})=
2\,{\cal F}_{ab}{^{\mu_1}}(x)\,h_{\mu_1\mu_2\alpha} g^{\alpha\beta}
h_{\mu_3\mu_4\beta}\,R^{(ax,\,bx)\mu_2 \mu_3\mu_4} }
\nonumber\\
&&+(2h_{ax\,bx\,\alpha}h_{\mu_1\mu_2\beta}g^{\alpha\beta} -
h_{\mu_2\alpha\,[\,ax} h_{bx\,]\,\mu_1\beta}g^{\alpha\beta})
R^{(ax,\,bx)\mu_1\mu_2}
\end{eqnarray}
where we have used (\ref{Rsim}) and (\ref{F2h}). Using
(\ref{F1hh}) and performing the integrations by parts
indicated by the longitudinal projectors we get
\begin{eqnarray}
\lefteqn{\textstyle{1\over12}
H_0\,(h_{\cdot \cdot \star}g^{\star \star}h_{\star \cdot
\cdot}R^{\cdot \cdot \cdot \cdot})=
2\{g_{\mu_1[\,ax}h_{bx\,]\,\mu_2 \mu_3}
-\epsilon_{abc}\phi^{\,cx}_{\,x_1}h_{\mu_1\mu_2\mu_3}
}\nonumber\\
&&+\epsilon_{abc}\phi^{\,cx}_{\,z}
(\phi^{\,dz}_{\,x_3}-\phi^{\,dz}_{\,x_2})g_{\mu_1 \,dz}g_{\mu_2\mu_3}
\} R^{(ax,\,bx) \mu_1 \mu_2 \mu_3}
\nonumber\\
&&-\{ 2h_{ax\,bx\,\alpha} g^{\alpha\beta}h_{\mu_3\mu_4\beta} -
h_{\mu_2\alpha\,[\,ax} h_{bx\,]\,\mu_1\beta}g^{\alpha\beta}
+2\epsilon_{abc}\phi^{\,cx}_{\,z}(\phi^{\,dz}_{\,x}-
\phi^{\,dz}_{\,x_1})h_{dz\,\mu_1\mu_2}
\nonumber\\
&&+2\epsilon_{abc}\phi^{\,cx}_{\,z}\phi^{\,dz}_{\,y}
(\phi^{\,ey}_{\,x}+\phi^{\,ey}_{\,x_1})g_{\mu_1\,dz}g_{\mu_2\,ey}
\}R^{(ax,\,bx)\mu_1\mu_2}
\label{Hr4}
\end{eqnarray}
We observe that in this expression does not appear contributions of the
type $R^{(ax,\,bx)\ldots\,cx\,\ldots}$. This is due to the fact that
there are no ``free"  two point propagators present in $h_{\cdot \cdot
\star}g^{\star \star}h_{\star \cdot \cdot}$. Also notice that the
rank five contribution cancels the term of rank five of
(\ref{Hr5}) as it was advanced in the preceding subsection.

What happens with the rank four of (\ref{Hr4})? We do not have at our
disposal a term of rank three in $J_3$ to continue the chain of
cancellations. This contribution has necessarily to vanish. This
condition follows from the relationship existing between the Hamiltonian
and the diffeomorphism in the extended loop representation. All the
results obtined heretofore for the Hamiltonian are valid for the
diffeomorphism with the replacement of the two-point-R by the
one-point-R. This means that the annulment of the rank four of
(\ref{Hr4}) is required from the diffeomorphism invariance of $J_3$
\footnote{There are not additional symmetry properties of the
one-point-R that allows to simplify the result more than in
(\ref{Hr4})}. One can see this requirement as a condition of consistency
for the analytical expression (\ref{j3}) to represent a diffeomorphism
invariant. The demonstration of this fact will be given in the appendix
III.

To conclude this subsection we resume: there is no remnant contribution
corresponding to the lower rank of $J_3$ and this term is responsible of
the closure (in a consistent way) of the chain of cancellations
associated with the action of the constraints in the extended loop
representation. This punctualization is also valid for the conventional
loop representation.

\section{Collecting the partial results}
The results of the preceding section show that it is possible to write
the following expression for the action of the Hamiltonian on the third
coefficient of the Jones polynomial
\begin{eqnarray}
\lefteqn{ \textstyle{1\over12}H_0(x)\,J_3 ({\bf R})=
2\epsilon_{abc}h_{\mu_1\mu_2\mu_3}R^{(ax,\,bx)\mu_1\,cx\,\mu_2 \mu_3} }
\nonumber\\
&&+2\epsilon_{abc}g_{\mu_1\mu_3}g_{\mu_2\mu_4}
[R^{(ax,\,bx)\mu_1 \,cx\, \mu_2  \mu_3 \mu_4}+
R^{(ax,\,bx)\mu_1 \mu_2 \,cx\, \mu_3 \mu_4}]
\nonumber\\
&&+\epsilon_{abc}g_{\mu_1\mu_4}g_{\mu_2\mu_3}
R^{(ax,\,bx)\mu_1\mu_2\,cx\,\mu_3\mu_4}
\label{result}
\end{eqnarray}
This expression admits several ulterior manipulations in the extended
representation: one that exploits the symmetry properties of the
propagators and the $R$'s under permutation of the indices and other
that points out towards a ``geometrical" \footnote{Remember that the
results of the extended representation can be always specialized to the
case of ordinary loops. This means that a geometrical interpretation can
in principle be possible for the analytical result (\ref{result}).}
interpretation of the result. As both are important and pursue different
goals  we will analize them for separate.

\subsection{Trying to annihilate $H_0\,J_3$}
At this point it is convenient to remember the initial motivation of the
calculation. From the general arguments of Sect. 2 it would be expected
the cancellation of (\ref{result}). But how can this expression be
zero?

The only general way to answer this question is by using symmetry
considerations. Notice that the symmetries {\it must work} in the case
of the one-point-R in order to make $J_3$ a solution of the
diffeomorphism constraint. The symmetry operations performed with
covariant indices include the properties of $\epsilon_{abc}$ and of the
propagators $g_{\cdot\cdot}$ and $h_{\cdot\cdot\cdot}$. The
contravariant indices can be moved using the overline operation (like
in (\ref{Rsim})) and the cyclicity.

The properties of the two-point-R under the overline of the indices were
totally exploited in the calculation. We have now to decompose these
combinations according to (\ref{rdosin}). For example, for the
contribution of rank six of (\ref{result}) one has
\begin{eqnarray}
\lefteqn{ R^{(ax,\,bx)\mu_1\,cx\,\mu_2 \mu_3}=
R^{(ax\,bx\,\mu_3\mu_2\,cx\,\mu_1)_c}
-R^{(ax\,\mu_1\,bx\,\mu_3\mu_2\,cx)_c} }
\nonumber\\
&&+R^{(ax\,\mu_1\,cx\,bx\,\mu_3\mu_2)_c}
-R^{(ax\,\mu_1\,cx\,\mu_2\,bx\,\mu_3)_c}
+R^{(ax\,\mu_1\,cx\,\mu_2\mu_3\,bx)_c}
\end{eqnarray}
This expression is reduced to the following using the cyclicity and
the symmetry properties of $h_{\cdot\cdot\cdot}$ and $\epsilon_{abc}$:
\begin{equation}
\epsilon_{abc}h_{\mu_1\mu_2\mu_3}
R^{(ax,\,bx)\mu_1\,cx\,\mu_2 \mu_3}=\epsilon_{abc}h_{\mu_1\mu_2\mu_3}
R^{(ax\,\mu_1 \,bx\, \mu_2 \,cx\,\mu_3)_c}
\end{equation}
Repeating the procedure for the others terms of (\ref{result}) one gets

\begin{eqnarray}
\lefteqn{ \textstyle{1\over12}H_0(x)\,J_3 ({\bf R})=
2\epsilon_{abc}h_{\mu_1\mu_2\mu_3}
\, R^{(ax\,\mu_1 \,bx\, \mu_2 \,cx\,\mu_3)_c} }
\nonumber\\
&&-6\epsilon_{abc}g_{\mu_1\mu_2}g_{\mu_3\mu_4}
[ R^{(ax\,\mu_1 \mu_3 \,bx\, \mu_2 \mu_4\,cx)_c}
+R^{(ax\,\mu_1 \mu_3 \,bx\, \mu_4 \mu_2\,cx)_c}
\nonumber\\
&&-R^{(ax\,\mu_1\,bx\,\mu_3\mu_2\,cx\,\mu_4)_c}]
\label{finalresult}
\end{eqnarray}
No further reduction is possible. This is the final result.

We see that the Hamiltonian does not annihilate the third coefficient of
the Jones polynomial for general multivector fields. As it was
emphasized before, the only difference between the Hamiltonian and the
diffeomorphism operators in the extended loop representation is to
change the two-point-R by the one-point-R. To end this subsection we
shall verify that the diffeomorphism operator annihilates $J_3$. From
(\ref{result}) one can write \footnote{The property (\ref{Rsim}) used to
simplfy the results for the hamiltonian corresponds to the following for
the diffeomorphism: $R^{(\overline{bx\,\mu_1\ldots\mu_n})_c}=(-1)^{n+1}
R^{(bx\,\mu_n\ldots\mu_1)_c}$.}
\begin{eqnarray}
\lefteqn{ \textstyle{1\over6}C_{ax}\,J_3 ({\bf R})=
2\epsilon_{abc}h_{\mu_1\mu_2\mu_3}R^{(bx\,\mu_1\,cx\,\mu_2 \mu_3)_c} }
\nonumber\\
&&+2\epsilon_{abc}g_{\mu_1\mu_3}g_{\mu_2\mu_4}
[R^{(bx\,\mu_1 \,cx\, \mu_2  \mu_3 \mu_4)_c}+
R^{(bx\,\mu_1 \mu_2 \,cx\, \mu_3 \mu_4)_c}]
\nonumber\\
&&+\epsilon_{abc}g_{\mu_1\mu_4}g_{\mu_2\mu_3}
R^{(bx\,\mu_1\mu_2\,cx\,\mu_3\mu_4)_c}
\label{resultdif}
\end{eqnarray}
This expression can be rewritten in the following way \footnote{The same
procedure in (\ref{finalresult}) generates antysimmetric expressions
with respect to the interchange of $bx$ with $cx$ and $ax$ with $cx$.}

\begin{eqnarray}
\lefteqn{\textstyle{1\over6} C_{ax}\,J_3 ({\bf R})=
\epsilon_{abc}h_{\mu_1\mu_2\mu_3}[R^{(bx\,\mu_1\,cx\,\mu_2 \mu_3)_c}
+R^{(cx\,\mu_1\,bx\,\mu_2 \mu_3)_c}] }
\nonumber\\
&&+\epsilon_{abc}g_{\mu_1\mu_3}g_{\mu_2\mu_4}
[R^{(bx\,\mu_1 \,cx\, \mu_2  \mu_3 \mu_4)_c}+
R^{(cx\,\mu_1 \,bx\, \mu_2  \mu_3 \mu_4)_c}
\nonumber\\
&&\hspace{2.6cm}+R^{(bx\,\mu_1 \mu_2 \,cx\, \mu_3 \mu_4)_c}
+R^{(cx\,\mu_1 \mu_2 \,bx\, \mu_3 \mu_4)_c}]
\nonumber\\
&&+\textstyle{1\over2}\epsilon_{abc}g_{\mu_1\mu_4}g_{\mu_2\mu_3}
[R^{(bx\,\mu_1\mu_2\,cx\,\mu_3\mu_4)_c}+
R^{(cx\,\mu_1\mu_2\,bx\,\mu_3\mu_4)_c}]\equiv 0
\label{finresultdif}
\end{eqnarray}
that vanishes identically in a formal sense.

\subsection{A geometrical interpretation for $H_0\,J_3$}
The connection between the extended loop and the loop representations
involves some technology that was introduced in \cite{DiGaGr2} and
developed in \cite{Gr} for the particular case of the third coefficient
of the Jones polynomial. In what follows we will incorporate the main
consequences of this procedure of reduction into the context of the
present discussion.

Perhaps the main virtue of the procedure of reduction developed in
\cite{Gr} is to show that new combinations of multivector fields of the
general form $R^{(ax,\,bx)\ldots\,cx\,\ldots}$ appear and that they have
a direct and simple geometrical interpretation when extended loops
are particularized to ordinary loops. The combinations are
\begin{equation}
R^{(ax,\,bx)\underline{cx}\,
\mu_1 \ldots \mu_n}:=\sum_{k=0}^{n}R^{(ax,\,bx)\, \mu_1 \ldots \mu_k\,cx\,
\mu_{k+1} \ldots \mu_n}
\label{underliner}
\end{equation}
and
\begin{equation}
R^{(ax,\,bx)\underline{\underline{cx}}\,
\mu_1 \ldots \mu_n}:=\sum_{k=0}^{n}R^{(ax,\,bx)\, \mu_1 \ldots \mu_k\,cx\,
\overline{\mu_{k+1} \ldots \mu_n}}
\label{uunderliner}
\end{equation}
The way that ordinary loops appear is simple: one has only to establish
that the multivector fields are now the ``multitangent fields"
$X^{\mu_1\ldots\mu_n}(\gamma)$ associated with the loop $\gamma$.
Besides the properties of the multivector fields that belong to
the general extended loop group, the multitangents have another
particularities like the algebraic constraint and the possibility to
express the loop as a composition of open paths. These
last properties are in the root of the procedure of reduction mentioned
above.

In order to gain simplicity we limit the discussion to a particular
case. Let $\gamma$ be a loop with a triple intersection at the point $x$
and denote $\gamma=\gamma_1\gamma_2\gamma_3$, being $\gamma_1$,
$\gamma_2$ and $\gamma_3$ the three ``petals" of the trefoil. It is
assumed that the origin $o$ of the loop lies in $\gamma_1$. A
multitangent field with three spatial points fixed at $x$ decomposes in
the following way
\begin{eqnarray}
\hspace{-0.4cm}
\lefteqn{ X^{\mu_1 \ldots \mu_i \, ax \,\mu_{i+1} \ldots
\mu_j\,bx\mu_{j+1}\ldots\mu_k\,cx\,\mu_{k+1}\ldots\mu_n}(\gamma) = }
\nonumber \\
&&\hspace{-0.4cm}
X^{\mu_1 \ldots \mu_i}(\gamma_1{^x_o}) \,T_1^{ax}\,
X^{\mu_{i+1} \ldots \mu_j}(\gamma_2)\,T_2^{bx}\,
X^{\mu_{j+1} \ldots \mu_k}(\gamma_3)\,T_3^{cx}\,
X^{\mu_{k+1} \ldots \mu_n}(\gamma_1{^o_x})
\label{tangent}
\end{eqnarray}
where $T^{ax}_m$ is the tangent at $x$  when one crosses the time
$m$ to this point and $\gamma^z_y$ indicates the portion of the loop
from $y$ to $z$. Using this result it is possible to show that \cite{Gr}
\begin{eqnarray}
\lefteqn{ \epsilon_{abc}
R^{(ax,\,bx)\underline{cx}\,\mu_1\ldots\mu_n}(\gamma)= }
\nonumber\\
&&\hspace{-0.4cm}-2\epsilon_{abc}T_1^{ax}T_2^{bx}T_3^{cx}
[R^{\mu_1\ldots\mu_n}(\gamma_1\gamma_2\overline{\gamma}_3)+
R^{\mu_1\ldots\mu_n}(\gamma_1\overline{\gamma}_2\gamma_3)+
R^{\mu_1\ldots\mu_n}(\overline{\gamma}_1\gamma_2\gamma_3)]
\nonumber\\
\label{rloop1}
\end{eqnarray}
and
\begin{eqnarray}
\lefteqn{ \hspace{-0.4cm}\epsilon_{abc}
R^{(ax,\,bx)\underline{\underline{cx}}\,\mu_1\ldots\mu_n}(\gamma)= }
\nonumber\\
&&\hspace{-0.4cm}-2\epsilon_{abc}T_1^{ax}T_2^{bx}T_3^{cx}
[R^{\mu_1\ldots\mu_n}(\gamma_2\gamma_1\overline{\gamma}_3)+
R^{\mu_1\ldots\mu_n}(\gamma_3\overline{\gamma}_2\gamma_1)+
R^{\mu_1\ldots\mu_n}(\overline{\gamma}_1\gamma_3\gamma_2)]
\nonumber\\
\label{rloop2}
\end{eqnarray}
{\it if} the set of indices $\mu_1\ldots\mu_n$ is cyclic (that is, if
$\mu_1\ldots\mu_n\,=\,(\mu_1\ldots\mu_n)_c$). The relevant fact here is
that the multitangents of rank $n$ have been {\it reconstructed} and
that they appear instead of a product of multitangent fields like in
(\ref{tangent}). The reconstruction is generated by the sum in $k$ that
defines the quantities ${\bf R}^{(ax,\,bx)\underline{cx}}$ and ${\bf
R}^{(ax,\,bx)\underline {\underline{cx}}}$. Supposse now that the greek
indices of the above expressions are contracted with appropriate
covariant indices $D_{\mu_1\ldots\mu_n}$ and let us perform the sum in
$n$ of the resulting expression. We get for the first case:

\begin{eqnarray}
\lefteqn{ \epsilon_{abc}\sum_{n=0}^{\infty} D_{\mu_1
\ldots \mu_n}\,R^{(ax,\,bx)\underline{cx}\,\mu_1\ldots\mu_n}(\gamma)= }
\nonumber\\
&&-2\epsilon_{abc}T_1^{ax}T_2^{bx}T_3^{cx}
[\psi(\gamma_1\gamma_2\overline{\gamma}_3)+
\psi(\gamma_1\overline{\gamma}_2\gamma_3)+
\psi(\overline{\gamma}_1\gamma_2\gamma_3)]
\label{nueva1}
\end{eqnarray}
with
\begin{equation}
\psi(\gamma)\equiv\psi[{\bf R}(\gamma)]=
\sum_{n=0}^{\infty} D_{\mu_1
\ldots \mu_n}\,R^{\mu_1\ldots\mu_n}(\gamma)
\end{equation}
and $\overline{\gamma}$ is the rerouted loop. If $\psi(\gamma)$ is a
knot invariant, then the left hand side of (\ref{nueva1}) acquires a
natural geometrical meaning. This is precisely the case of the result
(\ref{result}). Notice that

\begin{eqnarray}
\lefteqn{ h_{\mu_1\mu_2\mu_3}[
R^{(ax,\,bx)\underline{cx}\,\mu_1 \mu_2 \mu_3}-
R^{(ax,\,bx)\underline{\underline{cx}}\mu_1 \mu_2 \mu_3}]= }
\nonumber\\
&&2h_{\mu_1\mu_2\mu_3}[ R^{(ax,\,bx)\mu_1 \,cx\, \mu_2 \mu_3}
-R^{(bx,\,ax)\mu_1\,cx\,\mu_2\mu_3}]
\end{eqnarray}
and
\begin{eqnarray}
\lefteqn{ g_{\mu_1 \mu_3}g_{\mu_2 \mu_4}
[R^{(ax,\,bx)\underline{cx}\,\mu_1 \mu_2 \mu_3\mu_4}-
R^{(ax,\,bx)\underline{\underline{cx}}\,\mu_1 \mu_2 \mu_3\mu_4}]= }
\nonumber\\
&&g_{\mu_1 \mu_3}g_{\mu_2 \mu_4}
[2R^{(ax,\,bx)\mu_1 \,cx\, \mu_2  \mu_3 \mu_4}
-2R^{(bx,\,ax)\mu_1 \,cx\, \mu_2  \mu_3 \mu_4}
\nonumber\\
&&+R^{(ax,\,bx)\mu_1 \mu_2 \,cx\, \mu_3 \mu_4}
-R^{(bx,\,ax)\mu_1 \mu_2 \,cx\, \mu_4 \mu_3}]
\end{eqnarray}
Using these results it is immediate to show that

\begin{eqnarray}
\lefteqn{ \textstyle{1\over6}
H_0\,J_3({\bf R})=\epsilon_{abc}h_{\mu_1 \mu_2 \mu_3}
[R^{(ax,\,bx)\underline{cx}\,\mu_1 \mu_2 \mu_3} -
R^{(ax,\,bx)\underline{\underline{cx}}\,\mu_1 \mu_2 \mu_3}] }
\nonumber\\
&&+\epsilon_{abc}g_{\mu_1 \mu_3}g_{\mu_2 \mu_4}
[R^{(ax,\,bx)\underline{cx}\,\mu_1 \mu_2 \mu_3\mu_4})-
R^{(ax,\,bx)\underline{\underline{cx}}\,\mu_1 \mu_2 \mu_3\mu_4}]
\nonumber\\
&&+3\epsilon_{abc}
g_{\mu_1 \mu_2}g_{\mu_3 \mu_4}
[R^{(ax,\,bx)\mu_1\mu_3\,cx\,\mu_2\mu_4}+
R^{(ax,\,bx)\mu_1\mu_3\,cx\,\mu_4\mu_2}]
\label{nueva2}
\end{eqnarray}
The analytical expression of the second coefficient of the Jones
polynomial is

\begin{equation}
J_2({\bf R})= -3\{h_{\mu_1\mu_2\mu_3}\,R^{\mu_1\mu_2\mu_3}+
g_{\mu_1\mu_3}g_{\mu_2\mu_4}\,R^{\mu_1\mu_2\mu_3\mu_4}\}
\end{equation}
so,
\begin{eqnarray}
\lefteqn{ H_0\,J_3({\bf R})=
2\epsilon_{abc}\{\,J_2[\,{\bf R}^{(ax,\,bx)\underline{\underline{cx}}}\,]-
J_2[\,{\bf R}^{(ax,\,bx)\underline{cx}}\,]\,\}  }
\nonumber\\
&&+18\epsilon_{abc}
g_{\mu_1 \mu_2}g_{\mu_3 \mu_4}
[R^{(ax,\,bx)\mu_1\mu_3\,cx\,\mu_2\mu_4}+
R^{(ax,\,bx)\mu_1\mu_3\,cx\,\mu_4\mu_2}]
\label{nueva3}
\end{eqnarray}
We see that the analytical expression of the $J_2$ diffeomorphism
invariant appears naturally in $H_0\,J_3$. The remaining contributions
generate Gauss-linking numbers when ordinary loops are introduced. In
effect, according the results of reference \cite{Gr} one finds the
following result for the case of the trefoil

\begin{eqnarray}
\lefteqn{ \textstyle{1\over12}H_0(x)\,J_3 (\gamma)= -\epsilon_{abc}
T_1^{ax}T_2^{bx}T_3^{cx}\{ \,J_2(\gamma_1\gamma_2\gamma_3)-
J_2(\gamma_2\gamma_1\gamma_3)  }
\nonumber\\
&&+2[\varphi_G(\gamma_1,\gamma_2)-\varphi_G(\gamma_1,\gamma_3)]^2
+2[\varphi_G(\gamma_2,\gamma_3)-\varphi_G(\gamma_2,\gamma_1)]^2
\nonumber\\
&&+2[\varphi_G(\gamma_3,\gamma_1)-\varphi_G(\gamma_3,\gamma_2)]^2 \}
\label{nueva6}
\end{eqnarray}
where
\begin{equation}
\varphi_G(\gamma_i,\gamma_j):=g_{\mu_1\mu_2}\,X^{\mu_1}(\gamma_i)\,
X^{\mu_2}(\gamma_j)
\label{nueva7}
\end{equation}
gives the linking number of the loops $\gamma_i$ and $\gamma_j$. The
geometrical content of (\ref{nueva6}) is quite nontrivial. This result
valid formally for ordinary loops follows from the properties of the
quantities (\ref{underliner}) and (\ref{uunderliner}) at the level of
extended loops.

The form of (\ref{nueva6}) suggests that the cancellation of the pairs
of $J_2$ and $\varphi_G$ invariants could take place for certain
particular topologies of the loop. The possibility that $J_3(\gamma)$
could represent a nonperturbative quantum state of gravity acquires then
a new significance. This possibility remembers now the basic property of
the ``smoothened loops" \cite{smoothened}, that is the annihilation of
the loop wavefunction by the Hamiltonian when the domain of definition
is restricted to loops without intersections. A first approach has not
revealed any immediate solution of this kind. This question is currently
under progress.

\section{The Kauffman bracket}
We consider now the Kauffman bracket. As it was shown in Sect. 2, the
relevance of the Kauffman bracket in quantum gravity is based on the
fact that the exponential of the Chern-Simons form is a solution of the
constraints in the connection representation. The properties of the
Wilson loops (which are essential for both the Kauffman bracket and the
loop transform) are in the root of the fact that

\begin{equation}
H_{\Lambda}\;(\mbox{\footnotesize{Kauffman bracket}})\,\stackrel{?}{=}\,0
\label{bra1}
\end{equation}
This observation points out to the following: the fact that the Kauffman
bracket is annihilated by $H_{\Lambda}$ is in concordance with the
coherence and simplicity of the (conventional) loop representation. Any
failure of the condition (\ref{bra1}) will immediately rebound on the
relationship between the connection and the loop representations through
the loop transform.

In what follows we verify (\ref{bra1}) to third order in the
cosmological constant. For this we need the results of the
preceding sections as well as those of \cite{Gr}. In reference \cite{Gr}
it was demonstrated by direct calculation that

\begin{eqnarray}
H_0\,(\varphi_G \, J_2) -det \,q (x)\, J_2&=&
2\epsilon_{abc}
\{
\,J_2[{\bf R}^{(ax,\,bx)\underline{\underline{cx}}}]-
J_2[{\bf R}^{(ax,\,bx)\underline{cx}}]\}
\nonumber\\
&&+9\epsilon_{abc}
g_{\mu_1 \mu_2}g_{\mu_3 \mu_4}
R^{(ax,\,bx)\underline{\mu_1\mu_2}\mu_3 \,cx\,\mu_4}
\label{bra2}
\end{eqnarray}
where the underline of the two indices in the second term of the r.h.s.
represents the following combination of multivector fields:

\begin{equation}
R^{\underline{\mu_1\mu_2}\mu_3\ldots\mu_n}:=
\sum_{k=2}^{n-1}\sum_{i=k}^{n}
R^{\mu_3\mu_4\ldots\mu_k\mu_1\mu_{k+1}\ldots\mu_i\mu_2\mu_{i+1}\ldots\mu_n}
\end{equation}
It is possible to prove without difficulty that

\begin{eqnarray}
\lefteqn{ \epsilon_{abc}g_{\mu_1 \mu_2}g_{\mu_3 \mu_4}
R^{(ax,\,bx)\underline{\mu_1\mu_2}\mu_3 \,cx\,\mu_4}=
2\epsilon_{abc}g_{\mu_1 \mu_2}g_{\mu_3 \mu_4}\times }
\nonumber\\
&&[R^{(ax,\,bx)\mu_1\,cx\,(\mu_2 \mu_3\mu_4)_c}+
R^{(ax,\,bx)\mu_1\mu_3\,cx\,\mu_2\mu_4}+
R^{(ax,\,bx)\mu_1\mu_3\,cx\,\mu_4\mu_2}]
\label{bra3}
\end{eqnarray}
and that

\begin{equation}
\epsilon_{abc}g_{\mu_1 \mu_2}g_{\mu_3 \mu_4}
R^{(ax,\,bx)\mu_1\,cx\,(\mu_2\mu_3\mu_4)_c}\equiv 0
\label{bra4}
\end{equation}
by symmetry considerations. Introducing (\ref{bra3}) and (\ref{bra4})
into (\ref{bra2}) and using (\ref{hk}) and (\ref{nueva3}) we obtain

\begin{equation}
H_{\Lambda} \, K^{(3)}_{\Lambda} =0
\label{bra5}
\end{equation}
This result reinforces the role of the (conventional) loop transform for
quantum gravity.

\section{Conclusions}
The result (\ref{bra5}) follows after a long chain of calculations in
the extended loop representation. As we have seen, the calculations are
highly nontrivial and require a methodology proper of extended loops.
Moreover, the intermediate result (\ref{result}) admits an interesting
geometrical interpretation for ordinary loops that by itself opens a new
question about the solutions of the vacuum Hamiltonian constraint. The
question is: Is it possible that a topological restriction of the domain
of definition of $J_3(\gamma)$ makes $H_0\,J_3(\gamma)=0$? \footnote{The
restriction of the domain of definition of the loop wavefunctions
introduces a new problem: the use of characteristic functions in the
loop space. This problem is not completely understood at present and it
is shared by the smoothened loops of reference \cite{smoothened}.} An
intriguing fact that points in the same direction is the following: in
general $J_3$ does not satisfy all the Mandelstam identities that are
required for the quantum states of gravity. However, it is possible to
demonstrate that the Mandelstam identities will be recovered by $J_3$
precisely for those loops that would make $H_0\,J_3(\gamma)=0$
\footnote{In reference \cite{Gr} it was shown that the topological
condition required for $H_0\,J_3(\gamma)=0$ (in particular, that
$J_2(\gamma_1\gamma_2\gamma_3)=J_2(\gamma_2\gamma_1\gamma_3)$ for the
case of the unknot trefoil) makes $J_3(\gamma)$ to satisfy all the
Mandelstam identities.}.

In the analysis, a special care was taken to point out the
particularities of the extended loop calculus. The existence of a
systematic for the operation of the constraints as well as the intimate
relationship existing between the diffeomorphism and the Hamiltonian
constraints in the extended loop framework were emphasized in several
opportunities. With respect to this point one can add the following: the
power of calculation of the extended loop representation allows to raise
several important questions about knot theory and quantum gravity. These
questions can be summarized as follows: 1) Is it possible to construct
in a systematic way analytical expressions of diffeomorphism invariants
using the two and three point propagators of the Chern-Simons theory? 2)
Are other propagators (besides $g$ and $h$) relevant for quantum
gravity? 3) Is it possible to check the Mandelstam identities of the
diffeomorphism invariants in a systematic way? 4) There exist analytical
expressions constructed in terms of $g$'s and $h$'s different to that of
the exponential of the Gauss self-linking and the Kauffman bracket that
are invariant under diffeomorphisms {\it and} that satisfy the
Mandelstam identities? This last point is specially relevant for quantum
gravity. These topics are currently under study.

\section*{Acknowledgments}
I want to thank Cayetano Di Bartolo, Rodolfo Gambini and Jorge Pullin
for many fruitful discussions.

\section*{Appendix I}
In this appendix we shall see that
\begin{equation}
\delta{^{\phantom{T} cx}_{\mbox{\scriptsize{T}}\phantom{ax}
\alpha_k}}\epsilon^{\alpha_k \alpha_l \alpha_m}=
\epsilon^{cx\, \alpha_l \alpha_m}+(\phi^{\,cx}_{\,z_l}
-\phi^{\,cx}_{\,z_m})g^{\alpha_l \alpha_m}
\label{a11}
\end{equation}
{}From (\ref{deltat}) we have
\begin{eqnarray}
\delta{^{\phantom{T} cx}_{\mbox{\scriptsize{T}}\phantom{ax}
\alpha_k}}\epsilon^{\alpha_k \alpha_l \alpha_m}&=&
\epsilon^{cx\, \alpha_l \alpha_m}-\phi^{\,cx}_{\,z_k,\,a_k}
\epsilon^{\alpha_k \alpha_l \alpha_m}
\nonumber\\
&=&\epsilon^{cx\, \alpha_l \alpha_m}+\phi^{\,cx}_{\,z_k}\,
\partial_{\alpha_k}\,\epsilon^{\alpha_k \alpha_l \alpha_m}
\label{a12}
\end{eqnarray}
The divergence of $\epsilon^{\alpha_k \alpha_l \alpha_m}$ with respect
to the $\alpha_k$-entry can be written in the following way

\begin{eqnarray}
&&\partial_{\alpha_k}\,\epsilon^{\alpha_k \alpha_l \alpha_m}=
\epsilon^{c_k c_l c_m}\,\partial_{c_kz_k}\,[\delta(z_k-z_l)\,
\delta(z_k-z_m)]
\nonumber\\
&&=
\delta(z_k-z_l)\,\epsilon^{c_l c_m c_k}\,\partial_{c_k}\,\delta(z_l-z_m)
-
\delta(z_k-z_m)\,\epsilon^{c_m c_l c_k}\,\partial_{c_k}\,\delta(z_m-z_l)
\nonumber\\
&&=\delta(z_k-z_l)\,g^{\alpha_l \alpha_m}
-
\delta(z_k-z_m)\,g^{\alpha_m \alpha_l}
\label{a13}
\end{eqnarray}
where in the last step we have used (\ref{invmetrica}). Introducing
(\ref{a13}) into (\ref{a12}) we get directly (\ref{a11}).

\section*{Appendix II}
Here we consider in detail the calculation of ${\cal F}_{ab}^{\cdot} (x)\,
h_{\cdot \cdot \star}g^{\star \star}h_{\star \cdot
\cdot} R^{\cdot \cdot \cdot \cdot }$. We have
\begin{eqnarray}
&&{\cal F}_{ab}^{\mu_1} (x)h_{\mu_1 \mu_2 \alpha}g^{\alpha\beta}
h_{\mu_3 \mu_4 \beta}=
{\cal F}_{ab}^{\mu_1} (x)g_{\mu_1 \alpha_1}\epsilon^{\alpha_1\alpha_2\alpha_3}
g_{\alpha_2\mu_2}g_{\alpha_3\alpha}g^{\alpha\beta}h_{\beta\mu_3 \mu_4}
\nonumber\\
&&=-\epsilon_{abc}\delta{^{\phantom{T} cx}_{\mbox{\scriptsize{T}}\phantom{ax}
\alpha_1}}\epsilon^{\alpha_1\alpha_2\alpha_3}
g_{\alpha_2\mu_2}g_{\alpha_3\alpha}g^{\alpha\beta}h_{\beta\mu_3 \mu_4}
\nonumber\\
&&=-g_{\mu_2 [\,ax} g_{\,bx]\,\alpha}g^{\alpha\beta}h_{\beta\mu_3\mu_4}
-
\epsilon_{abc}(\phi^{\,cx}_{\,z_2}
-\phi^{\,cx}_{\,z_3})g^{\alpha_2 \alpha_3}
g_{\alpha_2\mu_2}g_{\alpha_3\alpha}g^{\alpha\beta}h_{\beta\mu_3 \mu_4}
\nonumber\\
&&=-g_{\mu_2 [\,ax} h_{\,bx]\,\mu_3\mu_4}
-\epsilon_{abc}\phi^{\,cx}_{\,z}(g^{dz \alpha_3}
g_{dz\,\mu_2}g_{\alpha_3\alpha}
-g^{\alpha_2\,dz}g_{\alpha_2\mu_2}
g_{dz\,\alpha})g^{\alpha\beta}h_{\beta\mu_3 \mu_4}
\nonumber\\
&&=-g_{\mu_2 [\,ax} h_{\,bx]\,\mu_3\mu_4}
-\epsilon_{abc}\phi^{\,cx}_{\,z}(
\delta{^{\phantom{T} dz}_{\mbox{\scriptsize{T}}\phantom{ax}\alpha}}
g_{dz\,\mu_2}
-
\delta{^{\phantom{T} dz}_{\mbox{\scriptsize{T}}\phantom{ax}\mu_2}}
g_{dz\,\alpha})g^{\alpha\beta}h_{\beta\mu_3 \mu_4}
\label{a21}
\end{eqnarray}
The $\mu_2$ index in the transverse projector indicates the appearance
of the first integration by parts (and the corresponding use of the
differential constraint) once the free indices are contracted with
contravariant indices of multivector fields. For the other projector one
has

\begin{eqnarray}
\lefteqn{
\delta{^{\phantom{T} dz}_{\mbox{\scriptsize{T}}\phantom{ax}\alpha}}
g^{\alpha\beta}h_{\beta\mu_3 \mu_4}=
\delta{^{\phantom{T} dz}_{\mbox{\scriptsize{T}}\phantom{ax}\alpha}}
\delta{^{\phantom{T} \alpha}_{\mbox{\scriptsize{T}}\phantom{ax}\pi_1}}
\epsilon^{\pi_1\pi_2\pi_3}
g_{\pi_2\mu_3}g_{\pi_3\mu_4} }
\nonumber\\
&&=
\{\epsilon^{dz\, \pi_1 \pi_2}+(\phi^{\,dz}_{\,y_2}
-\phi^{\,dz}_{\,y_3})g^{\pi_2 \pi_3}\}
g_{\pi_2\mu_3}g_{\pi_3\mu_4}
\nonumber\\
&&=
\epsilon^{dz\, \pi_2 \pi_3}g_{\pi_2\mu_3}g_{\pi_3\mu_4}
+\phi^{\,dz}_{\,y}
(\delta{^{\phantom{T} ey}_{\mbox{\scriptsize{T}}\phantom{ax}\mu_4}}
g_{ey\,\mu_3}
-
\delta{^{\phantom{T} ey}_{\mbox{\scriptsize{T}}\phantom{ax}\mu_3}}
g_{ey\,\mu_4})
\label{a22}
\end{eqnarray}
The chain stops at this point. Introducing (\ref{a22}) into (\ref{a21})
we get after a few direct manipulations the result (\ref{F1hh}).

\section*{Appendix III}
In this appendix we shall demonstrate that the quantity
\begin{eqnarray}
\lefteqn{ {\cal I}_{ax\,bx\,\mu_1\mu_2}:=
2h_{ax\,bx\,\alpha} g^{\alpha\beta}h_{\mu_1\mu_2\beta} -
h_{\mu_2\alpha\,[\,ax} h_{bx\,]\,\mu_1\beta}g^{\alpha\beta} }
\nonumber\\
&&+2\epsilon_{abc}\phi^{\,cx}_{\,z}(\phi^{\,dz}_{\,x}-
\phi^{\,dz}_{\,x_1})h_{dz\,\mu_1\mu_2}
+2\epsilon_{abc}\phi^{\,cx}_{\,z}\phi^{\,dz}_{\,y}
\phi^{\,ey}_{\,x_1}g_{\mu_1\,dz}g_{\mu_2\,ey}
\nonumber\\
&&+\epsilon_{abc}\phi^{\,cx}_{\,z}\phi^{\,dz}_{\,y}
\phi^{\,ey}_{\,x}(g_{\mu_1\,dz}g_{\mu_2\,ey}-g_{\mu_1\,ey}g_{\mu_2\,dz})
\label{a31}
\end{eqnarray}
vanishes identically (in a formal sense). In the above expression, $x$
is the point where the Hamiltonian is applied, $x_1$ is the spatial part
of $\mu_1$ and  $z$ and $y$ are continuous variables integrated in
${\cal R}^3$. We start by considering the following decompositon of the
three point propagator:

\begin{equation}
h_{ax\,\mu\alpha}=
\epsilon^{e_1e_2e_3}\epsilon_{a e_1 k}\phi^{\,kx}_{\,y}
g_{\mu\,e_2y}g_{\alpha e_3y}
= \phi^{\,ey}_{\,x}(g_{\mu\,ey}g_{\alpha ay}-g_{\mu\,ay}g_{\alpha ey})
\label{a32}
\end{equation}
When $\mu\equiv bx$ the above expession is reduced to

\begin{equation}
h_{ax\,bx\,\alpha}=\epsilon_{abc}\phi^{\,cy}_{\,x}\phi^{\,ey}_{\,x}
g_{\alpha ey}
\label{enap3}
\end{equation}
Using this fact the first term of ${\cal I}_{ax\,bx\,\mu_1\mu_2}$ can be
written in the following way

\begin{equation}
h_{ax\,bx\,\alpha} g^{\alpha\beta}h_{\mu_1\mu_2\beta}=
-\epsilon_{abc}\phi^{\,cx}_{\,z}\phi^{\,dz}_{\,x}
h_{dz\,\mu_1\mu_2}
\label{a33}
\end{equation}
The next step is to develop the second contribution of (\ref{a31}) in
terms of $g$'s. We have

\begin{eqnarray}
h_{\mu_2\alpha\,[\,ax} h_{bx\,]\,\mu_1\beta}g^{\alpha\beta}
&=&h_{\alpha\,ax\,[\mu_1} h_{\mu_2\,]\,bx\,\beta}g^{\alpha\beta}
\nonumber\\
&=&
\epsilon^{d_1d_2d_3}\epsilon^{e_1e_2e_3}g_{d_3z\,[\,\mu_1}g_{\mu_2\,]\,e_1y}
g_{ax\,d_2z}g_{bx\,e_2y}g_{d_1z\,e_3y}
\label{a34}
\end{eqnarray}
The $b$ index can be moved using the following identity
\begin{equation}
\epsilon^{e_1e_2e_3}g_{bx\,e_2y}=
\delta^{e_3}_b\,\phi^{\,e_1y}_{\,x}-\delta^{e_1}_b\,\phi^{\,e_3y}_{\,z}
\label{a35}
\end{equation}
We get from (\ref{a34})

\begin{eqnarray}
\lefteqn{ h_{\mu_2\alpha\,[\,ax} h_{bx\,]\,\mu_1\beta}g^{\alpha\beta} }
\nonumber\\
&&=\epsilon^{d_1d_2d_3}g_{ax\,d_2z}\{
g_{d_3z\,[\,\mu_1}g_{\mu_2\,]\,ey}g_{d_1z\,by}-
g_{d_3z\,[\,\mu_1}g_{\mu_2\,]\,by}g_{d_1z\,ey}
\}\phi^{\,ey}_{\,x}
\nonumber\\
&&=-h_{ey\,ax\,[\,\mu_1}g_{\mu_2\,]\,by}\phi^{\,ey}_{\,x}
+\epsilon^{d_1d_2d_3}g_{d_1z\,by}g_{ax\,d_2z}
g_{d_3z\,[\,\mu_1}g_{\mu_2\,]\,ey}
\phi^{\,ey}_{\,x}
\label{a36}
\end{eqnarray}
Developing now $\epsilon^{d_1d_2d_3}g_{d_1z\,by}$ according to
(\ref{a35}) we obtain from (\ref{a36}) a term with the $a$ and $b$
indices attached to the same $g$:

\begin{eqnarray}
 h_{\mu_2\alpha\,[\,ax} h_{bx\,]\,\mu_1\beta}g^{\alpha\beta}
&=&-h_{ey\,ax\,[\,\mu_1}g_{\mu_2\,]\,by}\phi^{\,ey}_{\,x}
+g_{ax\,bz}g_{dz\,[\,\mu_1}g_{\mu_2\,]\,ey}
\phi^{\,dz}_{\,y}\phi^{\,ey}_{\,x}
\nonumber\\
&&-g_{ax\,dz}g_{bz\,[\,\mu_1}g_{\mu_2\,]\,ey}
\phi^{\,dz}_{\,y}\phi^{\,ey}_{\,x}
\label{a37}
\end{eqnarray}
Notice that using (\ref{a32}) we can write

\begin{equation}
h_{ey\,ax\,[\,\mu_1}g_{\mu_2\,]\,by}\phi^{\,ey}_{\,x}
=
g_{az\,[\,\mu_1}g_{\mu_2\,]\,by}g_{dz\,ey}\phi^{\,dz}_{\,x}\phi^{\,ey}_{\,x}
-
g_{dz\,[\,\mu_1}g_{\mu_2\,]\,by}g_{dz\,ay}\phi^{\,dz}_{\,x}\phi^{\,ey}_{\,x}
\label{a38}
\end{equation}
and that

\begin{equation}
g_{dz\,ey}\phi^{\,dz}_{\,x}\phi^{\,ey}_{\,x}\equiv 0
\label{a39}
\end{equation}
Introducing then (\ref{a38}) into (\ref{a37}) one gets

\begin{equation}
h_{\mu_2\alpha\,[\,ax} h_{bx\,]\,\mu_1\beta}g^{\alpha\beta}
=
\epsilon_{abc}\phi^{\,cx}_{\,z}g_{dz\,[\,\mu_1}g_{\mu_2\,]\,ey}
\phi^{\,dz}_{\,y}\phi^{\,ey}_{\,x}
\label{a310}
\end{equation}
Applying the same procedure to the third term of (\ref{a31}) we obtain

\begin{eqnarray}
\epsilon_{abc}\phi^{\,cx}_{\,z}\phi^{\,dz}_{\,x_1}h_{dz\,\mu_1\mu_2}
&=&
\epsilon_{abc}\phi^{\,cx}_{\,z}\phi^{\,dz}_{\,x_1}
\epsilon^{e_1e_2e_3}g_{e_1y\,dz}g_{e_2y\,\mu_1}g_{e_3y\,\mu_2}
\nonumber\\
&=&
\epsilon_{abc}\phi^{\,cx}_{\,z}\phi^{\,dz}_{\,x_1}\phi^{\,ey}_{\,x_1}
(g_{dz\,ey}g_{a_1y\,\mu_2}-g_{dz\,a_1y}g_{e_y\,\mu_2})
\nonumber\\
&=&
\epsilon_{abc}\phi^{\,cx}_{\,z}\phi^{\,dz}_{\,y}\phi^{\,ey}_{\,x_1}
g_{dz\,\mu_1}g_{e_y\,\mu_2}
\label{a311}
\end{eqnarray}
where in the last step we used the following identity

\begin{equation}
\phi^{\,dz}_{\,x_1}g_{dz\,a_1y} \equiv -\phi^{\,dz}_{\,y}g_{dz\,a_1x_1}
\label{312}
\end{equation}
Introducing now (\ref{a33}), (\ref{a310}) and (\ref{a311}) into (\ref{a31})
we get

\begin{equation}
{\cal I}_{ax\,bx\,\mu_1\mu_2}=0
\label{a313}
\end{equation}

\end{document}